\documentclass[prb,twocolumn,showpacs,amsmath,amssymb]{revtex4}

\usepackage{graphicx}
\usepackage{dcolumn}
\usepackage{bm}

\begin{document}

\preprint{APS/123-QED}

\title{Anomalous temperature-dependent transport in YbNi$_2$B$_2$C\\ and its correlation to microstructural features}

\author{M. A. Avila$^{1,2}$}
\author{Y. Q. Wu$^{1,3}$}%
\author{C. L. Condron$^{1,4}$}%
\author{S. L. Bud'ko$^1$}%
\author{M. Kramer$^{1,3}$}%
\author{G. J. Miller$^{1,4}$}%
\author{P. C. Canfield$^{1,2}$}%
\affiliation{%
$^1$Ames Laboratory, $^2$Department of Physics and Astronomy,
$^3$Department of Materials Science and Engineering,
$^4$Department of Chemistry, Iowa State University, Ames, IA 50011
}%

\date{\today}

\begin{abstract}
We address the nature of the ligandal disorder leading to local
redistributions of Kondo temperatures, manifested as
annealing-induced changes in the transport behavior of the heavy
fermion system YbNi$_2$B$_2$C. The anomalous transport behavior
was fully characterized by temperature dependent resistivity
measurements in an extended range of $0.4<T<1000$~K for as-grown
and optimally annealed single crystals, and microstructural
changes between these two types of samples were investigated by
single-crystal x-ray diffraction and transmission electron
microscopy. Our results point to lattice dislocations as the most
likely candidate to be affecting the surrounding Yb ions, leading
to a distribution of Kondo temperatures. This effect combined with
the ability to control defect density with annealing offers the
possibility of further understanding of the more general problem
of the enhanced sensitivity of hybridized Kondo states to
disorder, particularly above the coherence temperature.

\end{abstract}



\maketitle

\section{\label{sec:intro}Introduction}

The transport properties of heavy fermion intermetallic systems
often display peculiar behaviors\cite{ste84b} whose origin remains
as yet unestablished and are subject of current interest. In these
systems the effective mass of conduction electrons is enhanced as
a result of hybridization between localized electronic orbitals
and delocalized conduction bands, and apparently these hybridized
states tend to be particularly sensitive to crystalline disorder,
manifesting strong sample-to-sample variations in the transport
properties\cite{ste84b,fra78a} which go beyond trivial differences
in residual resistivity.

The quaternary compound YbNi$_2$B$_2$C is a ytterbium-based heavy
fermion\cite{yat96a,dha96a} with an electronic specific heat
coefficient, $\gamma\approx500$~mJ/mol~K$^2$, and a Kondo
temperature, $T_K\approx10$~K: a temperature scale that is
conveniently isolated from other characteristic temperatures such
as superconducting or magnetic condensates ($T_c, T_N < 0.03$~K,
if they exist at all), and crystal electric field splitting
($T_{CEF}\approx100$~K)\cite{yat96a,dha96a,gra96a,ram00a,boot03a},
and therefore favorable for the study of Kondo physics.
Furthermore, an investigation of the effects of annealing on the
resistive behavior of boro-carbide single crystals\cite{mia02a}
found that YbNi$_2$B$_2$C displayed radical changes in
temperature-dependent resistance below room temperature, pointing
to the possibility that the transport properties of this compound
could be ``tuned'' to a certain extent by annealing, and thus
offer a model system to study the relationship between disorder
and hybridized states near the Fermi level.

In a previous report\cite{avi02a}, we demonstrated a significant
variation in the temperature dependent electrical resistivity and
thermoelectric power between as-grown crystals and crystals that
had undergone annealing at $950^{\circ}$C, whereas the
thermodynamic properties (heat capacity and magnetic
susceptibility) remained almost unchanged. We interpreted these
results in terms of redistributions of local Kondo temperatures
associated with ligandal disorder for a small (on the order of
1\%) fraction of the Yb sites. This hypothesis left open two
obvious questions: (i) What happens to the electrical resistivity
for temperatures greater than the maximum possible $T_K$ in this
proposed distribution? and (ii) What is the microscopic origin of
this distribution of Kondo temperatures?

In the present work, we first extend the temperature range of
resistivity measurements both higher (up to 1000~K) and lower
(down to 0.4~K), in order to fully characterize the transport
behavior of as-grown and optimally annealed crystals and provide
further evidence of the proposed model. We then present x-ray
single-crystal refinements and transmission electron microscopy
experiments, which provide new information on the types of
disorder that may be causing the anomalous Kondo temperatures in
the surrounding Yb sites.

\section{Experimental Details}

Single crystals of YbNi$_2$B$_2$C were grown from Ni$_2$B flux at
high temperatures as described elsewhere\cite{can98a,yat96a}. For
the resistivity measurements, a particularly clean and well-formed
single crystal plate measuring approximately
$7\times3\times0.2$~mm$^3$ was selected, polished on both surfaces
to remove most of the attached flux, and cut using a wire saw into
flat bars of typical dimensions $2\times0.5\times0.13$~mm$^3$ with
the length along the [100] direction. Electrical contacts were
placed on three unannealed samples in standard 4-probe linear
geometry, using Pt wires attached to a sample surface with
Ablebond 88-1 silver epoxy. For each slab, the sample weight and
dimensions were carefully measured and an evaluation of the sample
densities ($8.3\pm 0.2$~g/cm$^3$) was used to estimate an upper
limit of $\pm10\%$ for the overall geometry-related uncertainty in
calculating resistivity $\rho=RA/d$, where $A$ is the
cross-sectional area and $d$ is the distance between voltage
leads.

Two measured samples were then selected to undergo annealing for
comparison. The Pt wires were removed and the silver epoxy was
polished off, leading to a small ($\sim10\%$) decrease in the
sample's original weight due to an intentional, slight
overpolishing. Since the polishing occurred on a single flat
surface parallel to the current direction, the reduction in the
sample cross-section was assumed to be the same as the one in
mass. The samples were then annealed in vacuum for 150 hours at
$950^{\circ}$C (details of the annealing procedure are described
in ref.~\onlinecite{avi02a}).

Electrical resistance measurements below room temperature were
performed on commercial Quantum Design PPMS systems, allowing
measurements down to 1.8~K, or down to 0.4~K if the $^3$He option
is installed. Above room temperature, measurements were performed
on a custom-built high temperature resistivity system (HTRS) which
can be mounted on the same system used for sample annealing under
vacuum ($\sim10^{-6}$~Torr). A quartz probe is inserted in the
system with four leads where the other end of the sample contacts
are attached using silver epoxy. Temperature is measured with two
independent thermometers: Pt resistance temperature detector (RTD)
and type-R thermocouple. A 1~mA current is applied on the sample
by a Keithley 220 current source and the voltage is read on a HP
34420A nanovoltmeter. The RTD current is applied by a LakeShore
120 current source and the voltage is read on a HP 34401A
multimeter. The thermocouple voltage is read by another HP 34420A
nanovoltmeter. All these instruments are GPIB interfaced with a
computer running a custom-made software for data acquisition. To
reduce noise and avoid diode or thermoelectric effects, five
readings are taken with the current in each direction, and the ten
absolute voltage values are averaged.

A typical HTRS experiment proceeds as follows. 1) The system is
heated to 450~K for about an hour in air to cure the contacts,
then cooled back to room temperature. 2) The vacuum system is
turned on and allowed to achieve a dynamic vacuum of
$10^{-5}$~Torr or better. 3) The furnace temperature is ramped at
2~K/min or slower up to about 600~K and back to room temperature,
while the software acquires data points every 2 min. This first
ramp is used to allow further stabilization of the contacts which
often undergo small changes when heated, marked by irreproducible
shifts in the resistance behavior. 4) The definitive dataset is
acquired by ramping at 2~K/min or slower up to about 1000~K and
back to room temperature, while the software acquires data points
every 2 min. For the same reason of contact stability and in order
to obtain a good match between the curves from both measurement
systems, the low temperature measurements in the PPMS were
performed only after the high temperature ones, and for each
sample the data presented in figures \ref{polished} and
\ref{annealed} is the cool down from 1000~K to 300~K in the HTRS,
followed by the cool down from 300~K to 1.8~K or 0.4~K in the
PPMS. A certain level of mismatch still remains at room
temperature, both for the absolute value of the measured
resistance and for its slope, but these differences are considered
small enough to be neglected in the comparative analysis we
present in this work.

For the crystallographic studies, room temperature X-ray
diffraction (XRD) data of as-grown and annealed YbNi$_2$B$_2$C
single crystals were collected using a Bruker APEX diffractometer
with Mo $K_{\alpha}$ radiation ($\lambda=0.71073$~\AA) and a
detector-to-crystal distance of 5.08~cm. Data were collected for
the full sphere and were harvested by collecting three sets of
frames with 0.3$^{\circ}$ scans in $\omega$ for an exposure time
of 10 seconds per frame. The range of $2\theta$ extended from
3.0$^{\circ}$ to 56.0$^{\circ}$. Data were corrected for Lorentz
and polarization effects; absorption corrections were based on
fitting a function to the empirical transmission surface as
sampled by multiple equivalent reflections. Unit cell parameters
were indexed by peaks obtained from 90 frames of reciprocal space
images and then refined using all observed diffraction peaks after
data integration. Together with systematic absences, the space
group $I4/mmm$ (N. 139) was selected for subsequent structural
analysis. The structure solution was obtained by direct methods
and refined by full-matrix least-squares refinement of $F_o^2$
using the SHELXTL 6.10 package\cite{shel00a}.

For the transmission electron microscopy (TEM) evaluations, the
as-grown and annealed single crystals of YbNi$_2$B$_2$C were
characterized using a Philips CM30 transmission electron
microscope operating at 300~keV. The single crystals were placed
in a mortar with denatured ethyl alcohol and ground to a fine
powder. A pipette was used to place a small drop of the suspension
onto a holey carbon grid. The suspension was air-dried before
insertion into the microscope. The crystalline structure of
fragments which were electron transparent was evaluated using
selected area electron diffraction pattern (SADP). The defect
density was measured on select areas of the bright-field images
where the thickness was relatively uniform.

\section{Resistivity Measurements}

The main motivation for extending the resistivity measurements to
high temperatures was based on the assumption that, if the excess
scattering seen at room temperature and below in unannealed
samples arises from the contribution of Yb sites with Kondo
temperatures extending up through room temperature, there should
be a temperature $T_K^{max}$ above which the samples no longer
display anomalous scattering, and the difference between the
resistivity of unannealed sample and the annealed sample becomes a
``simple'' Matthiessen impurity term.

\begin{figure}[htb]
\includegraphics[angle=0,width=80mm]{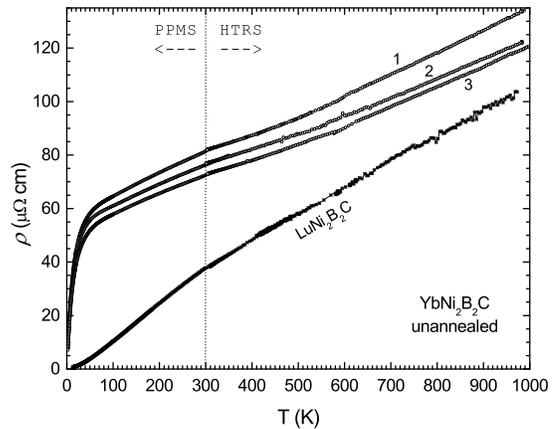}
\caption{\label{polished} Temperature dependence of the electrical
resistivity of three unannealed pieces of YbNi$_2$B$_2$C cut from
the same crystal. The differences between these curves are within
an uncertainty of 10\% in estimating the sample geometry. A
crystal of LuNi$_2$B$_2$C was also measured as a reference for the
conventional electron-phonon contribution to scattering.}
\end{figure}

The temperature dependence of the electrical resistivity for three
unannealed pieces of YbNi$_2$B$_2$C cut from the same crystal is
shown in Fig.~\ref{polished}. The general behavior below room
temperature is the same as has been previously reported for
as-grown samples of this
compound,\cite{mia02a,avi02a,bud97a,hos97a,yat99a,rat99a} i.e., a
high-scattering metallic behavior down to about 30~K, below which
the resistivity decreases rapidly. If normalized to their
respective resistance values at any given temperature, these three
curves essentially collapse into a single one in the entire
measured temperature interval, meaning that the selected crystal
was quite homogeneous and the differences seen in the three
measurements result from the overall $\pm10$\% uncertainty in
estimating the geometrical factor $A/d$ for each piece. The
residual resistivity ratio (defined as $R(300)/R(1.8)$) is
$RRR=10$ for all three samples, somewhat higher than $RRR=6-7$
found for several other unannealed crystals grown similarly
\cite{mia02a,avi02a}, indicating that by measuring the high
temperature region first we are already annealing out a small
portion of the disorder, but still maintaining most of the
characteristic unannealed sample behavior.

A crystal of LuNi$_2$B$_2$C was also measured to provide a
reference for the non-magnetic (e.g. standard electron-phonon)
contribution to scattering down to 16~K, below which it becomes
superconducting. The high temperature region is somewhat noisier
for this sample because its cross-section is about 3$\times$
larger than the others and therefore the actual measured
resistance is proportionally smaller. On the other hand, the
uncertainty in this sample's geometrical factor is equally
smaller, so its slope becomes a good reference. It is interesting
to note that, assuming $\rho_0l=412~\mu\Omega$~cm~\AA~for
LuNi$_2$B$_2$C,\cite{che98a} near 1000~K the measured resistivity
of order $100~\mu\Omega$~cm implies a mean free path $l\sim4$~\AA,
already comparable to the unit cell lattice parameter $a$ (the
Ioffe-Regel criterion), however the resistivity behavior is still
quite linear with a slope $\Delta\rho/\Delta T=0.1~\mu\Omega$~cm/K
between 600~K and 900~K. Above 600~K, all three YbNi$_2$B$_2$C
samples change slope and roughly follow that of the LuNi$_2$B$_2$C
sample (except for the nearly constant extra magnetic
contribution).

\begin{figure}[htb]
\includegraphics[angle=0,width=80mm]{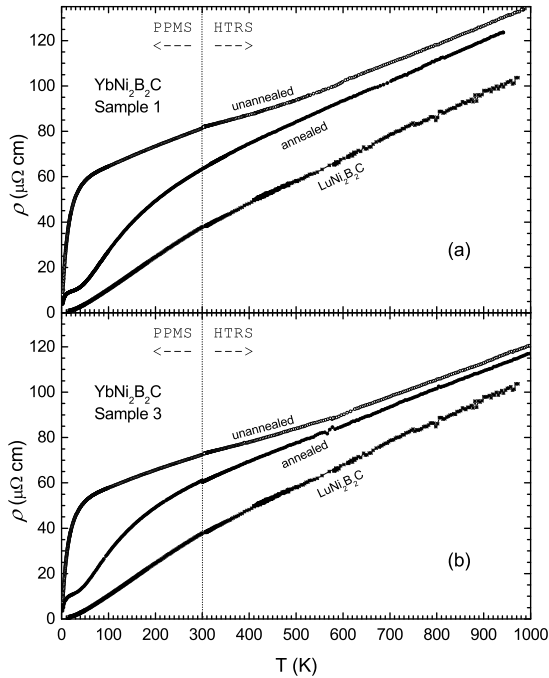}
\caption{\label{annealed} Temperature dependence of the electrical
resistivity of two pieces of YbNi$_2$B$_2$C cut from the same
crystal, before and after annealing at 950$^{\circ}$C for 150
hours. The measurement on LuNi$_2$B$_2$C serves as a reference for
the standard electron-phonon contribution to scattering.}
\end{figure}

Samples 1 and 3 were then annealed for 150 hours at 950$^{\circ}$C
and remeasured. Figures~\ref{annealed}a and \ref{annealed}b
present the comparison between unannealed and annealed resistivity
for each sample. Let us go through the comparison in detail for
sample 1. At high temperatures (above about 600~K), the curves for
the unannealed and annealed conditions of the sample run
essentially parallel to each other (and roughly parallel to
LuNi$_2$B$_2$C), meaning that in this region there is simply a
temperature-independent resistivity factor $\rho_0$, in accordance
with Matthiessen's rule. As we cool below 600~K the unannealed
curve maintains a more slowly decreasing resistivity (or,
conversely, a $relatively$ increased level of scattering), when
compared to the other two which continue to run parallel down to
about 200~K. Around this temperature the annealed YbNi$_2$B$_2$C
curve presents a broad shoulder, most likely related to the
thermal depopulation of the CEF multiplet with
$T_{cef}\approx100$~K, and therefore approaches the LuNi$_2$B$_2$C
curve. Between 50~K and 10~K the resistivity of the annealed
sample resembles a Kondo-minimum type behavior, and below 10~K its
resistivity once again drops rapidly due to the onset of coherent
scattering of the Yb ions, making its final approach towards the
conventional scattering level of the normal-state, non-magnetic
LuNi$_2$B$_2$C. Qualitatively similar results are found for sample
3. For both YbNi$_2$B$_2$C samples $RRR$ increased to 17 with
annealing, consistent with the previous results from annealed
crystals\cite{avi02a}.

Following the encouraging results above, we remeasured these
samples in the PPMS with the $^3$He option in order to extend the
resistivity curves down to our lowest measurable temperatures, so
as to check how annealing affects the behavior in this regime and
verify whether or not they follow the theoretical predictions for
a Fermi-liquid.

\begin{figure}[htb]
\includegraphics[angle=0,width=88mm]{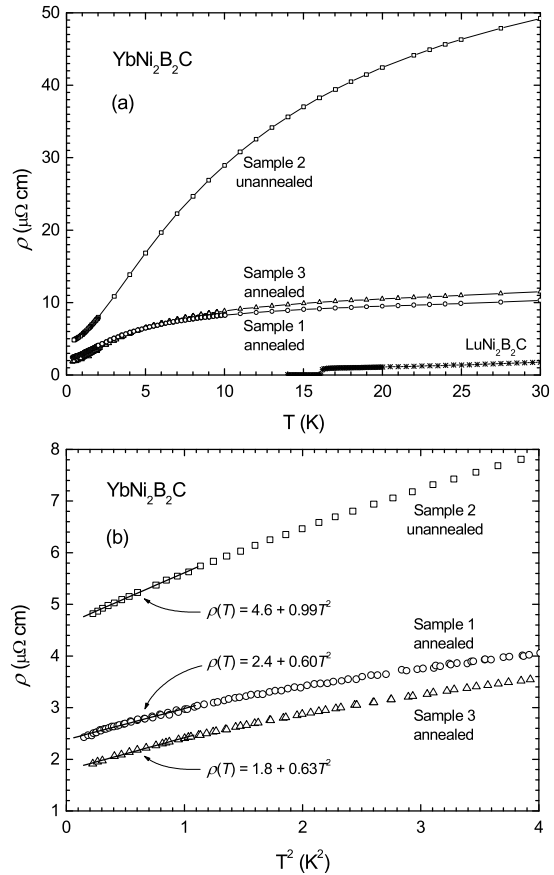}
\caption{\label{he3res} (a) Low temperature region of resistivity
for unannealed (Sample 2) and annealed (Samples 1 and 3)
YbNi$_2$B$_2$C. LuNi$_2$B$_2$C serves as comparison until it
becomes superconducting at 16~K. (b) $T^2$ dependence of the
resistivity for the YbNi$_2$B$_2$C samples at the lowest measured
temperatures.}
\end{figure}

In figure~\ref{he3res}a we show the temperature dependence of
resistivity below 30~K of sample 2 which was left unannealed, and
samples 1 and 3 (annealed). The decrease in resistivity below 10~K
is quite evident in the annealed samples, whereas the resistivity
of the unannealed one is already decreasing rapidly below 30~K. In
figure~\ref{he3res}b we plot the same data in the region below 2~K
as a function of $T^2$, and all three curves are slightly
sub-linear in this plot, indicating that they cannot be completely
described by a Fermi-liquid type relation $\rho(T)=\rho_0+AT^2$ in
this temperature range. In the graph we have included a fit of
this expression to the three datasets below 1~K. It seems that
with the decrease in disorder there is a decrease in both $\rho_0$
and $A$. The only other investigations of this type previously
reported on an unannealed crystal\cite{yat97a,yat99a} resulted in
$\rho(T)=12+1.2T^2$~$\mu\Omega$~cm, and are consistent with this
trend. Such a large reduction in $A$ is not expected within the
framework of the Fermi-liquid models, since our previous
study\cite{avi02a} found that the electronic specific heat
coefficient $\gamma$ changes very little with annealing (no more
than 10\% for the lowest measured temperatures). Even taking into
account this small change in $\gamma$, the ratio $A/\gamma_0^2$
still decreases from $0.4\times10^{-5}$ to
$0.3\times10^{-5}~\mu\Omega$~cm~(mol~K/mJ)$^2$ with annealing.
These values are smaller than the Kadowaki-Woods
ratio\cite{kado86a} of
$1\times10^{-5}~\mu\Omega$~cm~(mol~K/mJ)$^2$ found empirically for
many Fermi-liquid systems.

The difference of 2-3~$\mu\Omega$~cm in $\rho_0$ between our
annealed and unannealed YbNi$_2$B$_2$C samples is similar to the
results obtained by annealing studies on non-hybridizing
TmNi$_2$B$_2$C\cite{mia02b} and also comparable to the results of
Lu(Ni$_{1-x}$Co$_x$)$_2$B$_2$C substitution studies\cite{che98a}
when $x\approx1\%$. These similarities suggest that the optimal
annealing of single crystals grown by the Ni$_2$B flux growth
method is essentially removing lattice imperfections on the order
of 1\%, and in the coherent scattering regime YbNi$_2$B$_2$C
behaves like all other members of the family. Furthermore, the
difference in resistivity between annealed and unannealed samples
above 600~K (Fig.~\ref{annealed}) was found to be
3-6~$\mu\Omega$~cm, which is remarkably similar as well
(considering that the separations between the resistivity curves
are strongly affected by geometrical uncertainties at these high
temperatures) and therefore once again points to the return of a
more conventional scattering regime at very high temperatures.

Given the consistency of the extended transport measurements with
the hypothesis that ligandal disorder in the as-grown samples is
leading to redistributions of Kondo temperatures for a small
fraction of the Yb sites, the question that arises naturally is:
what are the types of disorder contributing to this behavior? To
address this question, we now present a series of structural and
microscopic experiments conducted on as-grown and annealed single
crystals.

\section{Single Crystal X-Ray Refinements}

\begin{figure}[htb]
\includegraphics[angle=0,width=40mm]{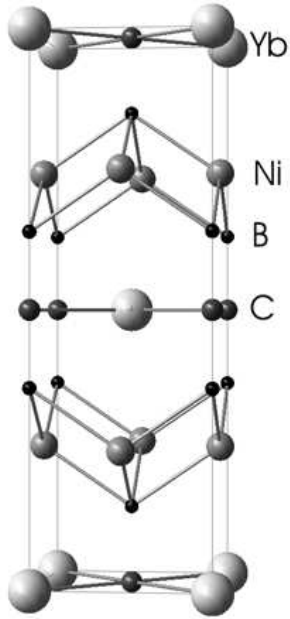}
\caption{\label{i4mmm} Unit cell diagram of YbNi$_2$B$_2$C.}
\end{figure}

The single-crystal XRD technique is useful to check for vacancies,
substitutions, superstructures, and other perturbations of
periodic atomic sites in the crystal. The structure refinement for
the as-grown and annealed YbNi$_2$B$_2$C was carried out under the
assumption that no structural disorder was present. By this
assumption all systematic absences pointed to the
expected\cite{sieg94b} crystal system (Tetragonal) and space group
($I4/mmm$) represented in figure~\ref{i4mmm}. Atomic positions
were assigned based on site symmetry and bond length and agree
with previously published single crystal data for other
$R$Ni$_2$B$_2$C series members\cite{sieg94a}. However, subtle
problems with the initial structure refinement of the as-grown
samples implied that the solution was not complete.

For most crystal structure refinements a parameter is added to
account for the accordance of extinction. In the SHELXTL software
the extinction parameter accounts for both primary and secondary
extinction\cite{shel00a}. Before the extinction parameter was
added to the refinement of the as-grown YbNi$_2$B$_2$C the R-index
was 0.1224 and the thermal parameters for Yb and B were not
convergent (table~\ref{tab:asgrown1}). After the extinction
parameter was added the refinement appeared to be stable with the
exception of the C thermal parameter (this is typical for light
elements in the presence of heavy elements). However, the
extinction parameter itself (0.24(2)) is larger than what is
normally expected (table~\ref{tab:asgrown2}). In solid state
structures a large extinction parameter usually indicates the
existence of a super-cell structure. Because super-cell structural
solutions were not found in this case, as indicated by the absence
of additional reflections, the large extinction parameter is most
likely due to the presence of dislocations. Refinement of the
annealed YbNi$_2$B$_2$C structure presented no problems
(extinction parameter 0.065(8)), suggesting that much of the
disorder present in the sample before the annealing process has
been removed. There is no evidence of vacancies or substitutions
within the resolution limit of this technique.

\begin{table}[htb]
\caption{\label{tab:asgrown1}As-grown YbNi$_2$B$_2$C without
extinction parameter. Space group $I4/mmm$, $a=3.479(5)$~\AA,
$c=10.617(2)$~\AA, $Z=2$, $R=0.1224$, $R_{\omega}=0.222$, 558
independent reflections, 66 observed ($I>2\sigma(I)$), 6
parameters.}
\begin{ruledtabular}
\begin{tabular}{cllll}
$Atom$ & $x$ & $y$ & $z$ & $U_{iso}$ \\
\hline
Yb & 0 & 0 & 0 & 0.00001 \\
Ni & 0 & 0.5 & 0.25 & 0.00258(1) \\
B & 0 & 0.5 & 0.360(7) & 0.00001 \\
C & 0.5 & 0.5 & 0 & 0.00917(1) \\
\end{tabular}
\end{ruledtabular}
\end{table}

\begin{table}[htb]
\caption{\label{tab:asgrown2}As-grown YbNi$_2$B$_2$C with
extinction parameter. Space group $I4/mmm$, $a=3.487(5)$~\AA,
$c=10.643(2)$~\AA, $Z=2$, $R=0.0205$, $R_{\omega}=0.0482$,
Extinction coefficient 0.24(2), 558 independent reflections, 66
observed ($I>2\sigma(I)$), 7 parameters.}
\begin{ruledtabular}
\begin{tabular}{cllll}
$Atom$ & $x$ & $y$ & $z$ & $U_{iso}$ \\
\hline
Yb & 0 & 0 & 0 & 0.0076(7) \\
Ni & 0 & 0.5 & 0.25 & 0.0066(6) \\
B & 0 & 0.5 & 0.360(7) & 0.0061(3) \\
C & 0.5 & 0.5 & 0 & 0.00001 \\
\end{tabular}
\end{ruledtabular}
\end{table}

\begin{table}[htb]
\caption{\label{tab:annealed}Annealed YbNi$_2$B$_2$C. Space group
$I4/mmm$, $a=3.487(5)$~\AA, $c=10.643(2)$~\AA, $Z=2$, $R=0.0266$,
$R_{\omega}=0.0612$, Extinction coefficient 0.065(8), 929
independent reflections, 67 observed ($I>2\sigma(I)$), 7
parameters.}
\begin{ruledtabular}
\begin{tabular}{cllll}
$Atom$ & $x$ & $y$ & $z$ & $U_{iso}$ \\
\hline
Yb & 0 & 0 & 0 & 0.0052(6) \\
Ni & 0 & 0.5 & 0.25 & 0.0042(6) \\
B & 0 & 0.5 & 0.361(2) & 0.0044(4) \\
C & 0.5 & 0.5 & 0 & 0.0065(4) \\
\end{tabular}
\end{ruledtabular}
\end{table}

\section{Transmission Electron Microscopy}

\begin{figure}[htb]
\includegraphics[angle=0,width=84mm]{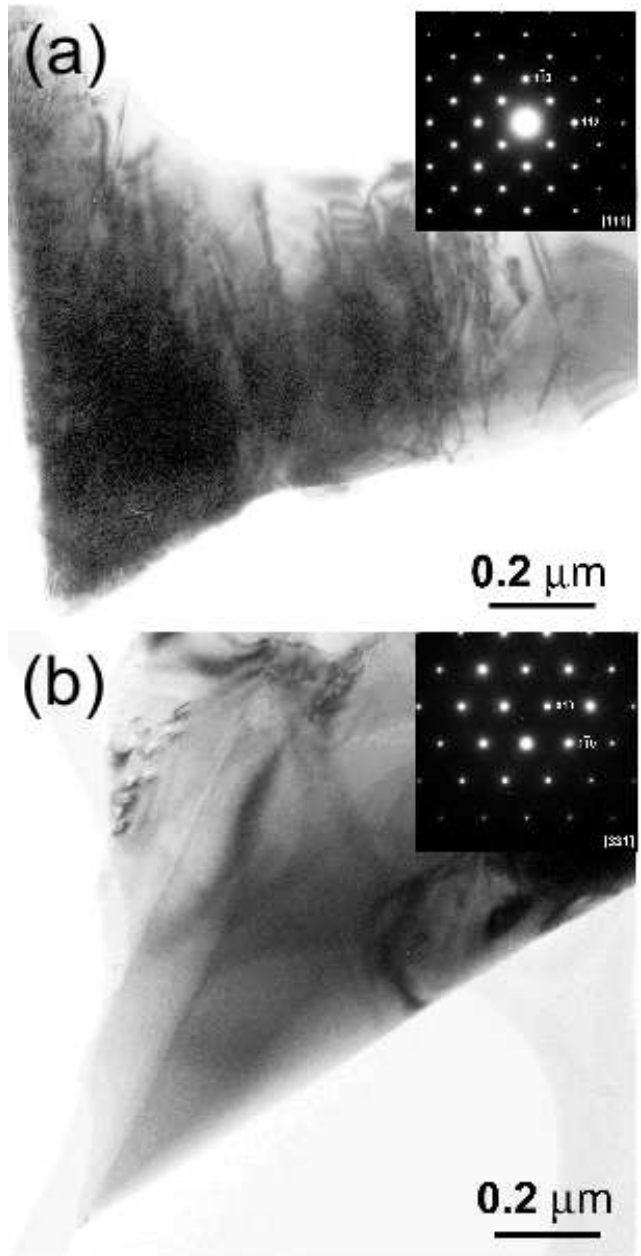}
\caption{\label{tem1} Bright-field TEM image of the crush-made
as-grown (a) and annealed (b) YbNi$_2$B$_2$C single grains. SAEDP
is shown as inset.}
\end{figure}

In order to obtain a more detailed and quantitative estimate of
these dislocations we studied the TEM patterns of crushed
crystals. Figure~\ref{tem1}(a) and (b) show two representative
bright field TEM images and the corresponding SADPs (insets) of
the crushed as-grown and annealed YbNi$_2$B$_2$C single grains,
respectively. Both SADPs are fully consistent with the known space
group ($I4/mmm$ (139)) for this compound, and they do have a
common reciprocal lattice vector, (110), which is approximately
equal for these two samples. There are no obvious second phases
and the diffraction spots are sharp, indicating well crystallized
samples. However, the as-grown crystallites show a surprisingly
large number of dislocations (Fig.~\ref{tem1}a). In contrast, the
annealed samples while exhibiting some regions of moderate defect
density, the defect density is qualitatively lower than the
as-grown sample. Using an estimated thickness of 800 and 500~nm
for the as-grown and annealed samples, we estimate a defect
density of $1.5\times10^8$ and $<6\times10^7$~cm$^{-2}$ for the
as-grown and annealed samples, respectively.

Although it is not expected that room temperature crushing of a
nominally brittle intermetallic would result in extensive
dislocation formation, it is not without precedence\cite{bera97a}.
The difference in the dislocation density between the as-grown and
annealed samples may be due to the presence of Frank-Read sources,
small defects which when subjected to high shear stresses are
capable of overcoming the Peierls forces resisting dislocation
mobility. This would imply that the as-grown crystals may have
many small, and hard to detect defects which are eliminated by
annealing.

\section{Discussion}

In our previous study\cite{avi02a} we made a semi-quantitative
evaluation of the fraction of Yb sites that would need to be
perturbed in order to account for the excess resistivity seen at
room temperature and below in the unannealed samples. Using the
expression for the increase in resistivity caused by a Kondo
impurity for temperatures below the Kondo temperature associated
with the impurity\cite{avi02a,tho85a}:

$\Delta\rho_{max}=\frac{\hbar}{e^2}\frac{4\pi c}{pk_F}(2l+1)$

where $c$ is the concentration of Yb ``impurities'', $k_{F}$ is
the Fermi momentum, $p$ is the number of electrons per atom, and
$l$ is the ytterbium $4f$ angular momentum, we estimated an
increase in the room temperature resistivity of about
$30~\mu\Omega cm$ for every 1\% concentration of affected Yb sites
with $T_K$ greater than 300~K. The extended resistivity
measurements and analysis performed in the present work have
confirmed and reinforced that initially proposed semi-quantitative
model. We have observed that when the sample is well into the
coherent scattering regime ($T\ll T_K\sim10$~K) or well above the
maximum perturbed Kondo temperature ($T\gg T_K^{max}\sim600$~K)
the Yb ions behave like the more conventional, non-hybridizing
rare-earth members in the $R$Ni$_2$B$_2$C series and the overall
resistivity decrease with annealing is of only 2-3~$\mu\Omega cm$,
once again consistent with a density of order 1\% of conventional
scattering sites.

The TEM experiments clearly showed the presence of defects in both
unannealed and annealed samples most likely associated with
lattice dislocations, and the defect density was estimated to be
\textit{at least} 2.5 times larger in the unannealed samples.
Similar qualitative changes were seen in the XRD data. With these
results we can propose a scenario where the dislocation sites act
as a source of strain fields which disturb nearby Yb sites with
intensity inversely proportional to their distance to the defect
origin, which would then lead to the appearance of a distribution
of local Kondo temperatures along the strain field decay.
Estimating the extension of a strain field around a dislocation
core is complicated, and will vary according to the nature of the
dislocation, the modulii of the material and the length of the
Burger's vector. However, we can once again attempt a
semi-quantitative analysis. Using the defect density of
$1.5\times10^8$~cm$^{-2}$ for the unannealed sample and assuming
that 1\% of the Yb sites are subject to a strain field, we can
estimate that all Yb sites within a radius of $460$~\AA~of each
dislocation core would be feeling the effect of the strain field.
This value corresponds to about 130 unit cells along the plane and
40 unit cells along the $c$-axis, and seems rather large given
that, according to Eshelby and others\cite{hir75a}, the region of
elastic limit should be on the order of 5 times the Burger's
vector. For our samples the dislocations appear to be of the
$1/2\langle110\rangle\{001\}$ type with corresponding Burger's
vector $b=a/2[110]\approx2.47$~\AA. We can still argue an enhanced
sensitivity of the hybridized Yb ions to the strain, but even so
it is likely that the extension of the strain field (or the
fraction of affected Yb sites) is being overestimated.

Whereas there are very clear changes in the microstructural data
(TEM and single-crystal XRD) with annealing, the link between
these changes and the changes in the transport measurements
remains only semi-quantitative.

\section{Conclusion}

In this work we have fully characterized the resistivity behavior
of annealed and unannealed YbNi$_2$B$_2$C single crystals in the
range of $0.4<T<1000$~K. Whereas the variability of resistivity
behavior for compounds with hybridizing moments has been long
suspected, our studies on YbNi$_2$B$_2$C finally allow a clear and
controlled demonstration of this effect. We were also able to use
single crystal XRD and TEM data to gain confidence in the claim
that the changes are indeed intrinsic to the compound which is
clearly single phase in both annealed and as-grown conditions. For
this particular material and growth process, lattice dislocations
seem to be the dominant defect type found in as-grown crystals,
and is most likely being responsible for environment changes in
the nearby Yb$^{3+}$ ions which lead to deviations of local Kondo
temperatures from their intrinsic value of $\sim10$~K. As a
consequence, it is important to realize that any attempt at
detailed analysis of transport properties such as those found for
YbNi$_2$B$_2$C (and these are not uncommon) are highly suspect,
and any comparison of experiments with theoretical models must
either take into account the sensitivity of the hybridizing
moments on the modelling side or guarantee a nearly defect-free
sample, a highly non-trivial experimental requirement.

\begin{acknowledgments}
We are thankful to F. Borsa for urging us to measure the
high-temperature of YbNi$_2$B$_2$C, to J. Schmalian for
stimulating discussions, to I. R. Fisher and T. Wiener for the
development of the high-temperature resistivity system, and to Y.
Mozharivskyj for assistance with the x-ray refinements. Ames
Laboratory is operated for the US Department of Energy by Iowa
State University under Contract No. W-7405-Eng-82. This work was
supported by the Director for Energy Research, Office of Basic
Energy Sciences.
\end{acknowledgments}


\begin{thebibliography}{24}
\expandafter\ifx\csname
natexlab\endcsname\relax\def\natexlab#1{#1}\fi
\expandafter\ifx\csname bibnamefont\endcsname\relax
  \def\bibnamefont#1{#1}\fi
\expandafter\ifx\csname bibfnamefont\endcsname\relax
  \def\bibfnamefont#1{#1}\fi
\expandafter\ifx\csname citenamefont\endcsname\relax
  \def\citenamefont#1{#1}\fi
\expandafter\ifx\csname url\endcsname\relax
  \def\url#1{\texttt{#1}}\fi
\expandafter\ifx\csname
urlprefix\endcsname\relax\def\urlprefix{URL }\fi
\providecommand{\bibinfo}[2]{#2}
\providecommand{\eprint}[2][]{\url{#2}}

\bibitem[{\citenamefont{Stewart}(1984)}]{ste84b}
\bibinfo{author}{\bibfnamefont{G.~R.} \bibnamefont{Stewart}},
  \bibinfo{journal}{Rev.\ Mod.\ Phys.} \textbf{\bibinfo{volume}{56}},
  \bibinfo{pages}{755} (\bibinfo{year}{1984}).

\bibitem[{\citenamefont{Franz et~al.}(1978)\citenamefont{Franz, Griessel,
  Steglich, and Wohlleben}}]{fra78a}
\bibinfo{author}{\bibfnamefont{W.}~\bibnamefont{Franz}},
  \bibinfo{author}{\bibfnamefont{A.}~\bibnamefont{Griessel}},
  \bibinfo{author}{\bibfnamefont{F.}~\bibnamefont{Steglich}}, \bibnamefont{and}
  \bibinfo{author}{\bibfnamefont{D.}~\bibnamefont{Wohlleben}},
  \bibinfo{journal}{Z.\ Phys.\ B} \textbf{\bibinfo{volume}{31}},
  \bibinfo{pages}{7} (\bibinfo{year}{1978}).

\bibitem[{\citenamefont{Yatskar et~al.}(1996)\citenamefont{Yatskar, Budraa,
  Beyermann, Canfield, and Bud'ko}}]{yat96a}
\bibinfo{author}{\bibfnamefont{A.}~\bibnamefont{Yatskar}},
  \bibinfo{author}{\bibfnamefont{N.~K.} \bibnamefont{Budraa}},
  \bibinfo{author}{\bibfnamefont{W.~P.} \bibnamefont{Beyermann}},
  \bibinfo{author}{\bibfnamefont{P.~C.} \bibnamefont{Canfield}},
  \bibnamefont{and} \bibinfo{author}{\bibfnamefont{S.~L.}
  \bibnamefont{Bud'ko}}, \bibinfo{journal}{Phys.\ Rev.\ B}
  \textbf{\bibinfo{volume}{54}}, \bibinfo{pages}{R3772} (\bibinfo{year}{1996}).

\bibitem[{\citenamefont{Dhar et~al.}(1996)\citenamefont{Dhar, Nagarajan,
  Hossain, Tominez, Godart, Gupta, and Vijayaraghavan}}]{dha96a}
\bibinfo{author}{\bibfnamefont{S.~K.} \bibnamefont{Dhar}},
  \bibinfo{author}{\bibfnamefont{R.}~\bibnamefont{Nagarajan}},
  \bibinfo{author}{\bibfnamefont{Z.}~\bibnamefont{Hossain}},
  \bibinfo{author}{\bibfnamefont{E.}~\bibnamefont{Tominez}},
  \bibinfo{author}{\bibfnamefont{C.}~\bibnamefont{Godart}},
  \bibinfo{author}{\bibfnamefont{L.~C.} \bibnamefont{Gupta}}, \bibnamefont{and}
  \bibinfo{author}{\bibfnamefont{R.}~\bibnamefont{Vijayaraghavan}},
  \bibinfo{journal}{Solid State Commun.} \textbf{\bibinfo{volume}{98}},
  \bibinfo{pages}{985} (\bibinfo{year}{1996}).

\bibitem[{\citenamefont{Grasser et~al.}(1996)\citenamefont{Grasser, Allenspach,
  Fauth, Henggeler, Mesot, Furrer, Rosenkranz, Vorderwisch, and
  Buchgeister}}]{gra96a}
\bibinfo{author}{\bibfnamefont{U.}~\bibnamefont{Grasser}},
  \bibinfo{author}{\bibfnamefont{P.}~\bibnamefont{Allenspach}},
  \bibinfo{author}{\bibfnamefont{F.}~\bibnamefont{Fauth}},
  \bibinfo{author}{\bibfnamefont{W.}~\bibnamefont{Henggeler}},
  \bibinfo{author}{\bibfnamefont{J.}~\bibnamefont{Mesot}},
  \bibinfo{author}{\bibfnamefont{A.}~\bibnamefont{Furrer}},
  \bibinfo{author}{\bibfnamefont{S.}~\bibnamefont{Rosenkranz}},
  \bibinfo{author}{\bibfnamefont{P.}~\bibnamefont{Vorderwisch}},
  \bibnamefont{and}
  \bibinfo{author}{\bibfnamefont{M.}~\bibnamefont{Buchgeister}},
  \bibinfo{journal}{Z.\ Phys.\ B} \textbf{\bibinfo{volume}{101}},
  \bibinfo{pages}{345} (\bibinfo{year}{1996}).

\bibitem[{\citenamefont{Rams et~al.}(2000)\citenamefont{Rams, Krolas, Bonville,
  Hodges, Hossain, Nagarajan, Dhar, Gupta, Alleno, and Godart}}]{ram00a}
\bibinfo{author}{\bibfnamefont{M.}~\bibnamefont{Rams}},
  \bibinfo{author}{\bibfnamefont{K.}~\bibnamefont{Krolas}},
  \bibinfo{author}{\bibfnamefont{P.}~\bibnamefont{Bonville}},
  \bibinfo{author}{\bibfnamefont{J.~A.} \bibnamefont{Hodges}},
  \bibinfo{author}{\bibfnamefont{Z.}~\bibnamefont{Hossain}},
  \bibinfo{author}{\bibfnamefont{R.}~\bibnamefont{Nagarajan}},
  \bibinfo{author}{\bibfnamefont{S.~K.} \bibnamefont{Dhar}},
  \bibinfo{author}{\bibfnamefont{L.~C.} \bibnamefont{Gupta}},
  \bibinfo{author}{\bibfnamefont{E.}~\bibnamefont{Alleno}}, \bibnamefont{and}
  \bibinfo{author}{\bibfnamefont{C.}~\bibnamefont{Godart}},
  \bibinfo{journal}{J.\ Magn.\ Magn.\ Mater.} \textbf{\bibinfo{volume}{219}},
  \bibinfo{pages}{15} (\bibinfo{year}{2000}).

\bibitem[{\citenamefont{Boothroyd et~al.}(2003)\citenamefont{Boothroyd,
  Barratt, Bonville, Canfield, Murani, Wildes, and Bewley}}]{boot03a}
\bibinfo{author}{\bibfnamefont{A.~T.} \bibnamefont{Boothroyd}},
  \bibinfo{author}{\bibfnamefont{J.~P.} \bibnamefont{Barratt}},
  \bibinfo{author}{\bibfnamefont{P.}~\bibnamefont{Bonville}},
  \bibinfo{author}{\bibfnamefont{P.~C.} \bibnamefont{Canfield}},
  \bibinfo{author}{\bibfnamefont{A.}~\bibnamefont{Murani}},
  \bibinfo{author}{\bibfnamefont{A.~R.} \bibnamefont{Wildes}},
  \bibnamefont{and} \bibinfo{author}{\bibfnamefont{R.~I.}
  \bibnamefont{Bewley}}, \bibinfo{journal}{Phys.\ Rev.\ B}
  \textbf{\bibinfo{volume}{67}}, \bibinfo{pages}{104407}
  (\bibinfo{year}{2003}).

\bibitem[{\citenamefont{Miao et~al.}(2002{\natexlab{a}})\citenamefont{Miao,
  Bud'ko, and Canfield}}]{mia02a}
\bibinfo{author}{\bibfnamefont{X.~Y.} \bibnamefont{Miao}},
  \bibinfo{author}{\bibfnamefont{S.~L.} \bibnamefont{Bud'ko}},
  \bibnamefont{and} \bibinfo{author}{\bibfnamefont{P.~C.}
  \bibnamefont{Canfield}}, \bibinfo{journal}{J. Alloys Compnds.}
  \textbf{\bibinfo{volume}{338}}, \bibinfo{pages}{13}
  (\bibinfo{year}{2002}{\natexlab{a}}).

\bibitem[{\citenamefont{Avila et~al.}(2002)\citenamefont{Avila, Bud'ko, and
  Canfield}}]{avi02a}
\bibinfo{author}{\bibfnamefont{M.~A.} \bibnamefont{Avila}},
  \bibinfo{author}{\bibfnamefont{S.~L.} \bibnamefont{Bud'ko}},
  \bibnamefont{and} \bibinfo{author}{\bibfnamefont{P.~C.}
  \bibnamefont{Canfield}}, \bibinfo{journal}{Phys.\ Rev.\ B}
  \textbf{\bibinfo{volume}{66}}, \bibinfo{pages}{132504}
  (\bibinfo{year}{2002}).

\bibitem[{\citenamefont{Canfield et~al.}(1998)\citenamefont{Canfield, Gammel,
  and Bishop}}]{can98a}
\bibinfo{author}{\bibfnamefont{P.~C.} \bibnamefont{Canfield}},
  \bibinfo{author}{\bibfnamefont{P.~L.} \bibnamefont{Gammel}},
  \bibnamefont{and} \bibinfo{author}{\bibfnamefont{D.~J.}
  \bibnamefont{Bishop}}, \bibinfo{journal}{Physics Today}
  \textbf{\bibinfo{volume}{51}}, \bibinfo{pages}{40} (\bibinfo{year}{1998}).

\bibitem[{\citenamefont{Sheldrick}(2000)}]{shel00a}
\bibinfo{author}{\bibfnamefont{G.~M.} \bibnamefont{Sheldrick}},
  \emph{\bibinfo{title}{SHELXTL, version 6.10}} (\bibinfo{publisher}{Bruker AXS
  Inc., Madison, WI}, \bibinfo{year}{2000}).

\bibitem[{\citenamefont{Bud'ko et~al.}(1997)\citenamefont{Bud'ko, Canfield,
  Yatskar, and Beyermann}}]{bud97a}
\bibinfo{author}{\bibfnamefont{S.~L.} \bibnamefont{Bud'ko}},
  \bibinfo{author}{\bibfnamefont{P.~C.} \bibnamefont{Canfield}},
  \bibinfo{author}{\bibfnamefont{A.}~\bibnamefont{Yatskar}}, \bibnamefont{and}
  \bibinfo{author}{\bibfnamefont{W.~P.} \bibnamefont{Beyermann}},
  \bibinfo{journal}{Physica B} \textbf{\bibinfo{volume}{230-232}},
  \bibinfo{pages}{859} (\bibinfo{year}{1997}).

\bibitem[{\citenamefont{Hossain et~al.}(1997)\citenamefont{Hossain, Nagarajan,
  Pattalwar, Dhar, Gupta, and Godart}}]{hos97a}
\bibinfo{author}{\bibfnamefont{Z.}~\bibnamefont{Hossain}},
  \bibinfo{author}{\bibfnamefont{R.}~\bibnamefont{Nagarajan}},
  \bibinfo{author}{\bibfnamefont{S.~M.} \bibnamefont{Pattalwar}},
  \bibinfo{author}{\bibfnamefont{S.~K.} \bibnamefont{Dhar}},
  \bibinfo{author}{\bibfnamefont{L.~C.} \bibnamefont{Gupta}}, \bibnamefont{and}
  \bibinfo{author}{\bibfnamefont{C.}~\bibnamefont{Godart}},
  \bibinfo{journal}{Physica B} \textbf{\bibinfo{volume}{230-232}},
  \bibinfo{pages}{865} (\bibinfo{year}{1997}).

\bibitem[{\citenamefont{Yatskar et~al.}(1999)\citenamefont{Yatskar, Mielke,
  Canfield, Lacerda, and Beyermann}}]{yat99a}
\bibinfo{author}{\bibfnamefont{A.}~\bibnamefont{Yatskar}},
  \bibinfo{author}{\bibfnamefont{C.~H.} \bibnamefont{Mielke}},
  \bibinfo{author}{\bibfnamefont{P.~C.} \bibnamefont{Canfield}},
  \bibinfo{author}{\bibfnamefont{A.~H.} \bibnamefont{Lacerda}},
  \bibnamefont{and} \bibinfo{author}{\bibfnamefont{W.~P.}
  \bibnamefont{Beyermann}}, \bibinfo{journal}{Phys.\ Rev.\ B}
  \textbf{\bibinfo{volume}{60}}, \bibinfo{pages}{8012} (\bibinfo{year}{1999}).

\bibitem[{\citenamefont{Rathnayaka et~al.}(1997)\citenamefont{Rathnayaka,
  Naugle, Lim, de~Andrade, Dickey, Amann, Maple, Bud'ko, Canfield, and
  Beyermann}}]{rat99a}
\bibinfo{author}{\bibfnamefont{K.~D.~D.} \bibnamefont{Rathnayaka}},
  \bibinfo{author}{\bibfnamefont{D.~G.} \bibnamefont{Naugle}},
  \bibinfo{author}{\bibfnamefont{S.}~\bibnamefont{Lim}},
  \bibinfo{author}{\bibfnamefont{M.~C.} \bibnamefont{de~Andrade}},
  \bibinfo{author}{\bibfnamefont{R.~P.} \bibnamefont{Dickey}},
  \bibinfo{author}{\bibfnamefont{A.}~\bibnamefont{Amann}},
  \bibinfo{author}{\bibfnamefont{M.~B.} \bibnamefont{Maple}},
  \bibinfo{author}{\bibfnamefont{S.~L.} \bibnamefont{Bud'ko}},
  \bibinfo{author}{\bibfnamefont{P.~C.} \bibnamefont{Canfield}},
  \bibnamefont{and} \bibinfo{author}{\bibfnamefont{W.~P.}
  \bibnamefont{Beyermann}}, \bibinfo{journal}{Int.\ J.\ Mod.\ Phys.}
  \textbf{\bibinfo{volume}{13}}, \bibinfo{pages}{3725} (\bibinfo{year}{1997}).

\bibitem[{\citenamefont{Cheon et~al.}(1998)\citenamefont{Cheon, Fisher, Kogan,
  Canfield, Miranovic, and Gammel}}]{che98a}
\bibinfo{author}{\bibfnamefont{K.~O.} \bibnamefont{Cheon}},
  \bibinfo{author}{\bibfnamefont{I.~R.} \bibnamefont{Fisher}},
  \bibinfo{author}{\bibfnamefont{V.~G.} \bibnamefont{Kogan}},
  \bibinfo{author}{\bibfnamefont{P.~C.} \bibnamefont{Canfield}},
  \bibinfo{author}{\bibfnamefont{P.}~\bibnamefont{Miranovic}},
  \bibnamefont{and} \bibinfo{author}{\bibfnamefont{P.~L.}
  \bibnamefont{Gammel}}, \bibinfo{journal}{Phys.\ Rev.\ B}
  \textbf{\bibinfo{volume}{58}}, \bibinfo{pages}{6463} (\bibinfo{year}{1998}).

\bibitem[{\citenamefont{Yatskar et~al.}(1997)\citenamefont{Yatskar, Bud'ko,
  Canfield, Beyermann, Schmiedeshoff, Torikachvili, Mielke, and
  Lacerda}}]{yat97a}
\bibinfo{author}{\bibfnamefont{A.}~\bibnamefont{Yatskar}},
  \bibinfo{author}{\bibfnamefont{S.~L.} \bibnamefont{Bud'ko}},
  \bibinfo{author}{\bibfnamefont{P.~C.} \bibnamefont{Canfield}},
  \bibinfo{author}{\bibfnamefont{W.~P.} \bibnamefont{Beyermann}},
  \bibinfo{author}{\bibfnamefont{G.~M.} \bibnamefont{Schmiedeshoff}},
  \bibinfo{author}{\bibfnamefont{M.~S.} \bibnamefont{Torikachvili}},
  \bibinfo{author}{\bibfnamefont{C.~H.} \bibnamefont{Mielke}},
  \bibnamefont{and} \bibinfo{author}{\bibfnamefont{A.}~\bibnamefont{Lacerda}},
  \bibinfo{journal}{Physica B} \textbf{\bibinfo{volume}{230-232}},
  \bibinfo{pages}{876} (\bibinfo{year}{1997}).

\bibitem[{\citenamefont{Kadowaki and Woods}(1986)}]{kado86a}
\bibinfo{author}{\bibfnamefont{K.}~\bibnamefont{Kadowaki}} \bibnamefont{and}
  \bibinfo{author}{\bibfnamefont{S.~B.} \bibnamefont{Woods}},
  \bibinfo{journal}{Solid State Commun.} \textbf{\bibinfo{volume}{58}},
  \bibinfo{pages}{507} (\bibinfo{year}{1986}).

\bibitem[{\citenamefont{Miao et~al.}(2002{\natexlab{b}})\citenamefont{Miao,
  Bud'ko, and Canfield}}]{mia02b}
\bibinfo{author}{\bibfnamefont{X.~Y.} \bibnamefont{Miao}},
  \bibinfo{author}{\bibfnamefont{S.~L.} \bibnamefont{Bud'ko}},
  \bibnamefont{and} \bibinfo{author}{\bibfnamefont{P.~C.}
  \bibnamefont{Canfield}}, \bibinfo{journal}{unpublished results related to
  ref.~\onlinecite{mia02a}}  (\bibinfo{year}{2002}{\natexlab{b}}).

\bibitem[{\citenamefont{Siegrist
  et~al.}(1994{\natexlab{a}})\citenamefont{Siegrist, Zandbergen, Cava,
  Krajewski, and Peck}}]{sieg94b}
\bibinfo{author}{\bibfnamefont{T.}~\bibnamefont{Siegrist}},
  \bibinfo{author}{\bibfnamefont{H.~W.} \bibnamefont{Zandbergen}},
  \bibinfo{author}{\bibfnamefont{R.~J.} \bibnamefont{Cava}},
  \bibinfo{author}{\bibfnamefont{J.~J.} \bibnamefont{Krajewski}},
  \bibnamefont{and} \bibinfo{author}{\bibfnamefont{W.~F.} \bibnamefont{Peck}},
  \bibinfo{journal}{Nature (London)} \textbf{\bibinfo{volume}{367}},
  \bibinfo{pages}{146} (\bibinfo{year}{1994}{\natexlab{a}}).

\bibitem[{\citenamefont{Siegrist
  et~al.}(1994{\natexlab{b}})\citenamefont{Siegrist, Cava, Krajewski, and
  Peck}}]{sieg94a}
\bibinfo{author}{\bibfnamefont{T.}~\bibnamefont{Siegrist}},
  \bibinfo{author}{\bibfnamefont{R.~J.} \bibnamefont{Cava}},
  \bibinfo{author}{\bibfnamefont{J.~J.} \bibnamefont{Krajewski}},
  \bibnamefont{and} \bibinfo{author}{\bibfnamefont{W.~F.} \bibnamefont{Peck}},
  \bibinfo{journal}{J.\ Alloys Compnds.} \textbf{\bibinfo{volume}{216}},
  \bibinfo{pages}{135} (\bibinfo{year}{1994}{\natexlab{b}}).

\bibitem[{\citenamefont{Beraha et~al.}(1997)\citenamefont{Beraha, Duneau,
  Klein, and Audier}}]{bera97a}
\bibinfo{author}{\bibfnamefont{L.}~\bibnamefont{Beraha}},
  \bibinfo{author}{\bibfnamefont{M.}~\bibnamefont{Duneau}},
  \bibinfo{author}{\bibfnamefont{H.}~\bibnamefont{Klein}}, \bibnamefont{and}
  \bibinfo{author}{\bibfnamefont{M.}~\bibnamefont{Audier}},
  \bibinfo{journal}{Phil.\ Mag.\ A} \textbf{\bibinfo{volume}{76}},
  \bibinfo{pages}{587} (\bibinfo{year}{1997}).

\bibitem[{\citenamefont{Thompson and Fisk}(1985)}]{tho85a}
\bibinfo{author}{\bibfnamefont{J.~D.} \bibnamefont{Thompson}} \bibnamefont{and}
  \bibinfo{author}{\bibfnamefont{Z.}~\bibnamefont{Fisk}},
  \bibinfo{journal}{Phys.\ Rev.\ B} \textbf{\bibinfo{volume}{31}},
  \bibinfo{pages}{389} (\bibinfo{year}{1985}).

\bibitem[{\citenamefont{Hirsch}(1975)}]{hir75a}
\bibinfo{author}{\bibfnamefont{P.~B.} \bibnamefont{Hirsch}},
  \emph{\bibinfo{title}{The Physics of Metals}} (\bibinfo{publisher}{Cambridge
  University Press}, \bibinfo{year}{1975}), vol. \bibinfo{volume}{2 Defects},
  p. \bibinfo{pages}{302}.

\end{thebibliography}

\end{document}